\documentclass[prb,aps,twocolumn,home,floats,showpacs]{revtex4}

\usepackage{graphics}
\usepackage{amsmath}
\usepackage{dcolumn}
\usepackage{citesort}
\begin{document}

\title {{\rm\small\hfill (submitted to Phys. Rev. B)}\\
First-principles kinetic Monte Carlo simulations for heterogeneous catalysis,\\
applied to the CO oxidation at RuO$_2$(110)}

\author{Karsten Reuter}
\affiliation{Fritz-Haber-Institut der Max-Planck-Gesellschaft, Faradayweg
4-6, D-14195 Berlin, Germany}

\author{Matthias Scheffler}
\affiliation{Fritz-Haber-Institut der Max-Planck-Gesellschaft, Faradayweg
4-6, D-14195 Berlin, Germany}

\received{4 October 2005}

\begin{abstract}
We describe a first-principles statistical mechanics approach enabling us to simulate the steady-state situation of heterogeneous catalysis. In a first step density-functional theory together with transition-state theory is employed to obtain the energetics of all relevant elementary processes. Subsequently the statistical mechanics problem is solved by the kinetic Monte Carlo method, which fully accounts for the correlations, fluctuations, and spatial distributions of the chemicals at the surface of the catalyst under steady-state conditions. Applying this approach to the catalytic oxidation of CO at RuO$_2$(110), we determine the surface atomic structure and composition in reactive environments ranging from ultra-high vacuum (UHV) to technologically relevant conditions, i.e. up to pressures of several atmospheres and elevated temperatures. We also compute the CO$_2$ formation rates (turnover frequencies). The results are in quantitative agreement with all existing experimental data. We find that the high catalytic activity of this system is intimately connected with a disordered, dynamic surface ``phase'' with significant compositional fluctuations. In this active state the catalytic function results from a self-regulating interplay of several elementary processes.
\end{abstract}



\maketitle

\section{Introduction}

Rational design and advancement in materials science will ultimately rely on an atomic-scale understanding of the targeted functionality. In macroscopic systems of technological relevance, this functionality typically results from the interplay of a large number of distinct atomic-scale processes. This makes quantitative calculations challenging, since it requires not only to compute many individual elementary processes with high accuracy to reach the aspired predictive power, but also to appropriately account for the statistical mechanics of the interplay of all these processes. Over the last years, modern multi-scale methods that suitably combine first-principles electronic structure calculations with thermodynamics and statistical mechanics have advanced sufficiently to make this goal feasible.\cite{reuter05a}

In heterogeneous catalysis the targeted functionality is the efficient conversion of a supply of chemicals $A$ and $B$ at the surface of a solid catalyst material into a product $C$ that is then transported away. Since this corresponds to a thermodynamic open system, even steady-state conditions, where the production of $C$ proceeds at a stable rate, are determined by kinetics. A quantitative computation of the steady-state rate, measured typically as a turnover frequency (TOF, number of product molecules per unit area and second), requires therefore to explicitly follow the time evolution of a large enough surface area, fully treating the interplay of all relevant underlying atomic-scale processes. For the example studied in this paper, the atomistic processes include the motion of the gas-phase molecules, dissociation, adsorption, surface diffusion, surface chemical reactions and desorption. Since many of these processes are activated and thus rare, macroscopic time scales need to be simulated, to arrive at meaningful conclusions concerning the effect of the statistical interplay. Thus, although individual processes occur on picosecond time scales, this may mean in practice to simulate the system over time scales up to the order of seconds or longer. 

An additional level of complexity is encountered when aspiring a predictive, quantitative modeling that is based on understanding, and that is applicable to realistic environmental situations of varying temperatures and pressures. This then not only excludes the use of empirical or fitted parameters, in which often several not further specified processes are effectively ``lumped together'', but also excludes the use of a mean-field description as e.g. microkinetic rate equation approaches \cite{masel}. As we will further qualify below, parameters with clear microscopic meaning and a full consideration of the correlations, fluctuations and spatial distributions of the chemicals at the catalyst surface are crucial to properly understand and potentially rationally design macroscopic catalytic functionality. Each individual elementary step must thus at first be treated separately, and then combined with all the others within an appropriate framework.

In a recent Letter we presented a first-principles statistical mechanics setup suitable to tackle this challenge \cite{reuter04}, and in this paper we give a detailed account of the employed methodology. We use density-functional theory (DFT) together with transition-state theory (TST) to accurately obtain the energetics of all relevant processes, and subsequently solve the statistical mechanics problem by kinetic Monte Carlo (kMC) simulations. This two-step approach enables us to gain microscopic insight into the system, following its full dynamics from picoseconds up to seconds. In particular, the approach is employed here to provide an {\em ab initio} description of the composition and structure of the catalyst surface in reactive environments ranging from ultra-high vacuum (UHV) to technologically relevant conditions with pressures of the order of atmospheres and room temperature or higher. We also quantitatively compute the catalyst activity in terms of TOFs under all of these conditions.

In the earlier Letter we used the CO oxidation at RuO$_2$(110) to illustrate the new quality and novel insights gained by such methodology, and we will continue to use this model system for this purpose here. Of course, this system is very interesting in its own right, too. Originally, there was work on supported catalysts by Cant, Hicks and Lennon \cite{cant78}, as well as on single crystals by Peden and Goodman \cite{peden86}, showing that what they called Ru catalysts exhibited high activity for CO oxidation under reactant pressures of the order of atmospheres. Extensive experimental and theoretical work has by now shown that in the corresponding reactive environments in fact RuO$_2$ forms at the surface \cite{boettcher97,boettcher99,over00,kim01f,reuter02,over03,assmann04}, and actuates the catalytic function \cite{kim01c,liu01,fan01,wang02}. Specifically, at the Ru(0001) model catalyst surface, it is an epitaxial RuO$_2$(110) film that is formed, and although domain boundaries and steps are present, their influence on the catalytic function is not significant.\cite{over03,he05} Our theoretical modeling will therefore focus on the surface processes at a RuO$_2$(110) facet under steady-state conditions. The theoretical results will be compared with existing experimental data \cite{peden86,wang02}, and with results obtained previously within the computationally much less demanding ``constrained thermodynamics'' approach \cite{reuter03a}. 

Having already briefly communicated the highlights of this work \cite{reuter04}, a large part of the present paper is devoted to a detailed description of the employed methodology. For this purpose we first establish in sections IIA and B the general theoretical framework for the combined DFT+kMC scheme, before addressing in section IIC the specific microscopic parameters calculated for the RuO$_2$(110) system. The method is then employed in section III to analyze the surface composition and structure, showing that high catalytic activity in this system is intimately connected with a disordered and dynamic ``phase'' at the surface, in which the observable catalytic function results from the self-regulating action of several different elementary processes.

\section{Theoretical framework}

\subsection{First-principles kinetic Monte Carlo simulations}

We aim at a first-principles statistical mechanics description, which provides atomic-scale understanding of the steady-state catalytic function by explicitly considering the detailed statistical interplay of all elementary processes, i.e. by fully accounting for correlations, fluctuations and spatial distributions.  This requires consideration of a large enough system size, but due to the kinetic nature of steady-state catalysis also to follow the evolution over time scales that are long enough to obtain reliable averages. Particularly the latter is quite demanding, since elementary processes in heterogeneous catalysis are typically activated and thus rare. While individual events occur on electronic to atomic time scales ($10^{-15} - 10^{-12}$\,sec), the time between consecutive events can be many orders of magnitude longer, possibly rendering it necessary to simulate up to seconds or more in order to arrive at meaningful conclusions about the effect of the statistical interplay.\cite{reuter05a}

Kinetic Monte Carlo (kMC) algorithms suitably address this problem by providing a numerical solution to the Markovian master equation, which describes the dynamic system evolution by efficiently coarse-graining the molecular dynamics to the decisive rare events and properly averaging over the irrelevant short-time dynamics.\cite{bortz75,gillespie76,voter86,kang89,fichthorn91} This allows to access time scales of the order of seconds or longer, even for mesoscopically-sized systems, while still retaining the full atomistic information.\cite{voter02} A kMC simulation proceeds by generating a sequence of system configurations. At each configuration, all possible elementary processes $p$ and the rates $r_p$ at which they occur are evaluated. Appropriately weighted by these different rates one of the possible processes is then executed randomly, leading to the next system configuration. In practice this is achieved by determining the total rate $R = \sum_{p=1}^P r_p$ summing over all $P$ possible processes, and executing process $k$, which fulfills the condition
\begin{equation}
\sum_{p=1}^k r_p \;\ge\; \rho_1 R \;\ge\; \sum_{p=1}^{k-1} r_p \quad,
\end{equation}
where $\rho_1 \in [0,1[$ is a random number. This way, the kMC algorithm effectively simulates stochastic processes described by a Poisson distribution, and a direct and unambiguous relationship to real time is established by advancing the clock by
\begin{equation}
t \;\rightarrow\; t - \frac{ln(\rho_2)}{R} \quad,
\end{equation}
where $\rho_2 \in [0,1[$ is another random number.\cite{fichthorn91}

The crucial ingredients to a kMC simulation are therefore the analysis and identification of all possibly relevant elementary processes at a given system configuration, and the determination of the associated rates. These rates are hitherto often either guessed or fitted such that the results of the kMC simulations match e.g. some existing experimental data. Modern first-principles kMC simulations \cite{ruggerone97,ovesson99,hansen00,fichthorn00,kratzer02} use instead rates obtained by electronic structure theory calculations, ensuring that the parameters fed into the kMC simulation have a clear microscopic meaning. To keep the number of elementary processes (and thus the number of required rates) tractable, it can be useful to suitably coarse-grain the system onto a lattice. This is the approach we will employ here, and for the present case of CO oxidation at RuO$_2$(110) this then leads to an exhaustive list of 26 processes. We note, however, that for more complex systems it may not be straightforward to identify a suitable lattice, or even if a lattice can be found, it may be difficult to identify and consider all of the possible processes occurring on it.\cite{reuter05a,henkelman03}

\subsection{Rates for elementary processes}

We now describe a consistent framework for the computations of the rates of elementary processes needed for a first-principles kMC simulation. Since the employed methodology is not restricted to the application to the RuO$_2$(110) model catalyst, we describe the framework in this section in more general terms, and refer only occasionally to the implementation in the RuO$_2$(110) system for illustration.

Let us consider a discretized lattice model of a surface, where each surface unit-cell of area $A_{uc}$ is represented by $st$ different site types (e.g. bridge and cus sites in the application to RuO$_2$(110) discussed below). This surface is exposed to a thermal gas composed of $i$ different species with masses $m_i$. At the accuracy level relevant for our study, this gas is well described by ideal gas laws, and is then sufficiently characterized by the temperature $T$ and partial pressures $p_i$. In the context of heterogeneous catalysis, relevant process types occurring in the system are adsorption and desorption processes between the gas phase and the lattice, surface chemical reactions, as well as diffusion processes between lattice sites.

\subsubsection{Adsorption and desorption: General}

The adsorption rate of species $i$ onto a free site of type $st$ depends on the kinetic impingement onto the whole surface unit-cell, as well as a local sticking coefficient $\tilde{S}_{st,i}(T)$, which governs which fraction of these impinging particles actually sticks to the given free site,
\begin{equation}
r^{\rm ad}_{st,i}(T,p_i) \;=\; \tilde{S}_{st,i}(T) \; \frac{p_i A_{uc}}{\sqrt{2\pi m_i k_B T}} \quad ,
\label{adrate}
\end{equation}
where $k_B$ is the Boltzmann constant. The impinging particles from the thermal gas have initially a randomly distributed lateral position of their center of gravity over the unit-cell, a random distribution over their possible internal degrees of freedom (e.g. molecular orientations) and Maxwell-Boltzmann distributed velocities. The local sticking coefficient represents thus a statistical average over these degrees of freedom of the impinging gas phase particles of species $i$, and gives only the fraction that sticks to a given free site of the specified site type $st$. 

If there is only one site type per surface unit-cell, $\tilde{S}_{st,i}(T)$ is identical to the often studied initial sticking coefficient, $S_{o,i}(T)$. \cite{darling95,gross98} The latter also describes a similar statistical average of sticking probabilities, but does not distinguish between site types. It thus represents the average sticking to all sites of a given clean surface and we have
\begin{equation}
S_{o,i}(T) \;=\; \sum_{st} \tilde{S}_{st,i}(T) \quad .
\end{equation}
If there are different adsorption site types in the surface unit-cell, knowledge of $S_{o,i}(T)$ is not sufficient for a kMC simulation. For the latter we need to know specifically the rate with which particles adsorb in each of the different available site types, and the relevant quantity is then $\tilde{S}_{st,i}(T)$.

In the so-called hole model for adsorption \cite{karikorpi87}, one considers that only gas phase particles of species $i$ with initial lateral position within a given area $A_{st,i}$ around the site $st$ have a chance to stick to the site. The local sticking coefficient will then contain a term $(\frac{A_{st,i}}{A_{uc}}) \le 1$, reducing the overall impingement onto the whole surface unit-cell contained in eq. (\ref{adrate}) to this active area. Note, that due to this term the overall adsorption rate $r^{\rm ad}_{st,i}(T,p_i)$ does also not depend on the actual choice of the surface unit-cell; only the gas phase particles impinging through the active area $A_{st,i}$ contribute to it. In this respect, the total sum of active areas can also not exceed the whole surface unit-cell area, since then more particles would be considered than the actual overall impingement rate. Hence, the choice of active areas is additionally limited by the condition
\begin{equation}
\sum_{st} A_{st,i} \;\le\; A_{uc} \quad .
\end{equation}

\begin{figure}[t!]
\scalebox{0.35}{\includegraphics{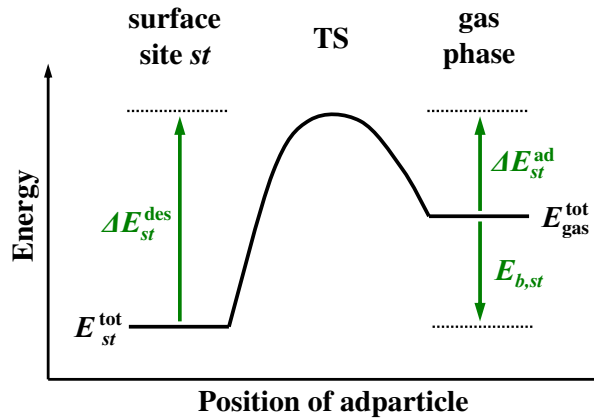}}
\caption{Potential-energy surface (schematic) along the minimum energy path connecting the free gas phase and adsorbed state in surface site $st$. Shown is a case, where the adsorption process is exothermic ($E_{b,st,i} < 0$), but activated ($\Delta E^{\rm ad}_{\rm st,i} > 0$).}
\label{mep}
\end{figure}

In a classical picture, sticking occurs if the particle's initial velocity is high enough to surmount the barriers along the trajectory to the surface. 
If all particles of species $i$ impinging within the active area $A_{st,i}$ around the site type $st$ had to pass the same maximum energetic barrier $\Delta E^{\rm ad}_{st,i} \ge 0$, the local sticking coefficient would simply be given by $(\frac{A_{st,i}}{A_{uc}}) \, {\rm exp} \left(- \frac{\Delta E^{\rm ad}_{st,i}}{k_B T}\right)$.\cite{karikorpi87} For a general form of the high-dimensional potential-energy surface (PES) seen by the impinging particles, this can be suitably generalized to
\begin{equation}
\tilde{S}_{st,i}(T) \;=\; f^{\rm ad}_{st,i}(T) \; \left( \frac{A_{st,i}}{A_{uc}} \right) \;
{\rm exp} \left(- \frac{\Delta E^{\rm ad}_{st,i}}{k_B T}\right) \quad .
\label{stick_general}
\end{equation}
Here, $\Delta E^{\rm ad}_{st,i}$ denotes the maximum barrier height along the minimum energy path (MEP) connecting the gas phase and adsorbed state in the PES as schematically shown in Fig. \ref{mep}. The factor $f^{\rm ad}_{st,i}(T) < 1$ accounts for a further reduction in the sticking probability, if particles with certain initial states (lateral positions over the unit-cell, internal degrees of freedom) are not efficiently steered along the MEP and are then possibly reflected by a higher barrier. 

Desorption of particles from occupied sites of the lattice represents the time-reversed process to adsorption, and as such has to fulfill detailed balance or microscopic reversibility. This principle ensures that the description of the free energy landscape for the forward and backward process between gas phase and adsorbed state is done consistently, and requires that the rate of adsorption and desorption are related by
\begin{eqnarray}
\nonumber
\frac{r^{\rm ad}_{st,i}(T,p_i)}{r^{\rm des}_{st,i}(T)} &=&
{\rm exp}\left(\frac{\Delta G_{st,i}(T,p_i)}{k_BT}\right) \\
&\approx& 
{\rm exp}\left(\frac{\mu_{{\rm gas},i}(T,p_i) - F_{st,i}(T)}{k_BT}\right) \quad ,
\label{equilibrium}
\end{eqnarray}
Here, $\Delta G_{st,i}(T,p_i)$ is the change in Gibbs free energy between the gas phase and the adsorbed state at site $st$. Neglecting the small $p_iV$ term for the adsorbed state \cite{reuter02a}, this is approximately given by the difference between the chemical potential $\mu_{{\rm gas},i}(T,p_i)$ of the particle in the gas phase and the free energy $F_{st,i}(T) = E^{\rm tot}_{st,i} - k_BT {\rm ln}(q^{\rm vib}_{st,i})$ in the adsorbed state. $E^{\rm tot}_{st,i}$ is hereby the total energy, and the partition function $q^{\rm vib}_{st,i}$ accounts for the vibrational degrees of freedom in the adsorbed state.

The chemical potential of an ideal gas can be written as sum of total energy 
$E^{\rm tot}_{{\rm gas},i}$ and free energy contributions from the translational and internal modes $\Delta \mu_{{\rm gas},i}(T,p_i)$, \cite{mcquarrie76}
\begin{eqnarray}
\nonumber
\mu_{{\rm gas},i}(T,p_i) &=& E^{\rm tot}_{{\rm gas},i} \;+\; \Delta \mu_{{\rm gas},i}(T,p_i) \\ \nonumber
                         &=& E^{\rm tot}_{{\rm gas},i} \;-\; k_BT {\rm ln}\left[ \left( \frac{2\pi m_ik_BT}{h^2} \right)^{3/2} \; \frac{k_BT}{p_i} \right] \\
&-& k_BT {\rm ln}(q^{\rm int}_{{\rm gas},i}) \quad ,
\label{mu_gas}
\end{eqnarray}
where $q^{\rm int}_{{\rm gas},i}$ is the partition function accounting for the internal degrees of freedom of the free gas phase particle. With the chosen sign convention, $\mu_{{\rm gas},i}(T,p_i)$ approaches thus $- \infty$ in the limit of an infinitely dilute gas. For diatomic gas molecules like O$_2$ and CO in the present case, $\Delta \mu_{{\rm gas},i}(T,p_i)$ can be readily calculated from first-principles, yielding results that are, at all temperatures and pressures of interest to us here, virtually indistinguishable from the experimental values listed in thermochemical tables.\cite{reuter02a} 

Introducing the binding energy for the particle $i$ at site $st$ and with respect to its reference state in the gas phase, $E_{b,st,i} = E^{\rm tot}_{st,i} - E^{\rm tot}_{{\rm gas},i}$ (with $E_{b,st,i} < 0$ for exothermicity), cf. Fig. \ref{mep}, allows to rewrite eq. (\ref{equilibrium}) to
\begin{eqnarray}
\nonumber
\frac{r^{\rm ad}_{st,i}(T,p_i)}{r^{\rm des}_{st,i}(T)} &=&
{\rm exp}\left(\frac{\Delta G_{st,i}(T,p_i)}{k_BT}\right) \\
&\approx& 
\frac{1}{q^{\rm vib}_{st,i}} \; {\rm exp}\left(\frac{\Delta \mu_{{\rm gas},i}(T,p_i) - E_{b,st,i}}{k_BT}\right) \quad .
\label{equilibrium2}
\end{eqnarray}
Equations (\ref{adrate}), (\ref{stick_general}), and (\ref{equilibrium2}) provide a consistent framework for the description of the kinetics of adsorption and corresponding desorption processes. The system-specific quantities required to fix the adsorption and desorption rates via these equations are $f^{\rm ad}_{st,i}(T)$, $A_{st,i}$ and $\Delta E^{\rm ad}_{st,i}$ describing the adsorption process, and $E_{b,st,i}$ and $q^{\rm vib}_{st,i}$ describing the bound state at the surface. The equations apply equally to processes that are unimolecular in either direction, or to processes that describe dissociative adsorption and in turn associative desorption. The processes can involve only one site of the surface lattice, as e.g. in the case of non-dissociative CO adsorption and desorption in the RuO$_2$(110) example discussed below. Or they can involve more sites, possibly due to dissociative adsorption or because a larger molecule blocks several sites upon adsorption. In the catalytic context, also surface chemical reactions involving direct desorption of the product may be described with the same framework of equations, viewing such processes then as dissociative adsorption and associative desorption of the product molecule.

\subsubsection{Adsorption and desorption: Transition state theory}

The computationally most demanding quantity entering the hitherto established framework is the factor $f^{\rm ad}_{st,i}(T)$, accounting for a reduction in the sticking probability due to those particles not efficiently steered along the MEP. An accurate calculation of $f^{\rm ad}_{st,i}(T)$ requires dynamical simulations of a statistically meaningful number of trajectories of impinging particles, for which information about large parts of the underlying high-dimensional PES is required.\cite{gross98} In the case of activated adsorption, an approximate $f^{\rm ad}_{st,i}(T)$ can also be determined through transition state theory (TST), relying on less PES information. As described above, $f^{\rm ad}_{st,i}(T)$ summarizes the dependence on all other degrees of freedom of the gas phase particle apart from the reaction coordinate along the MEP. In TST this is approximated by the fraction of correspondingly accessible states for thermalized particles at the MEP barrier and in the initial gas phase, i.e.
\begin{equation}
f^{\rm ad}_{st,i}(T) \;\approx\; f^{\rm ad, TST}_{st,i}(T) \;=\;
\frac{q^{\rm vib}_{{\rm TS}(st,i)}}{q^{\rm trans,2D}_{{\rm gas},i} q^{\rm int}_{{\rm gas},i}} \quad .
\label{tst_ad}
\end{equation}
Here, $q^{\rm vib}_{{\rm TS}(st,i)}$ is the partition function at the barrier (i.e. the transition state, TS), and we have separated the total partition function in the initial state conveniently into contributions from the remaining two lateral translational degrees of freedom $q^{\rm trans,2D}_{{\rm gas},i} = A_{\rm uc} \frac{2\pi m_ik_BT}{h^2}$ over the surface unit-cell, and the internal degrees of freedom $q^{\rm int}_{{\rm gas},i}$ of the free gas phase particle. The information about the PES required for an evaluation of $f^{\rm ad, TST}_{st,i}(T)$ is thus the location of the TS, as well as the local PES around it.

Equation (\ref{tst_ad}) introduces an approximation, since in contrast to the unspecified $f^{\rm ad}_{st,i}(T)$ the fraction of partition functions accounts no longer for possible recrossings of the impinging particles after they have successfully surmounted the MEP barrier, and also limits the accessible states at the TS to those of an equilibrated particle. Together with the initially made assumption of a classical barrier, these are the three classic assumptions underlying TST \cite{laidler87}. Indeed, using $f^{\rm ad, TST}_{st,i}(T)$ in the adsorption rate in the detailed balance equation, eq. (\ref{equilibrium2}), yields an expression for the desorption rate familiar from TST
\begin{equation}
r^{\rm des}_{st,i}(T) \;=\; f^{\rm des,TST}_{st,i}(T) \; \left(\frac{k_BT}{h}\right)
\; {\rm exp} \left(- \frac{\Delta E^{\rm des}_{st,i}}{k_B T}\right) \quad ,
\label{desrate}
\end{equation}
where
\begin{equation}
f^{\rm des,TST}_{st,i}(T) \;=\; \frac{\left(\frac{A_{st,i}}{A_{uc}} \; q^{\rm vib}_{{\rm TS}(st,i)} \right)}{q^{\rm vib}_{st,i}} \quad ,
\label{f_destst}
\end{equation}
and we have realized that along the MEP, the adsorption barrier $\Delta E^{\rm ad}_{st,i}$ and desorption barrier $\Delta E^{\rm des}_{st,i}$ are related through, cf. Fig. \ref{mep},
\begin{equation}
\Delta E^{\rm des}_{st,i} \;=\; \Delta E^{\rm ad}_{st,i} \;-\; E_{b,st,i} \quad .
\label{mep_relation}
\end{equation}

\subsubsection{Diffusion}

\begin{figure*}[t!]
\scalebox{0.70}{\includegraphics{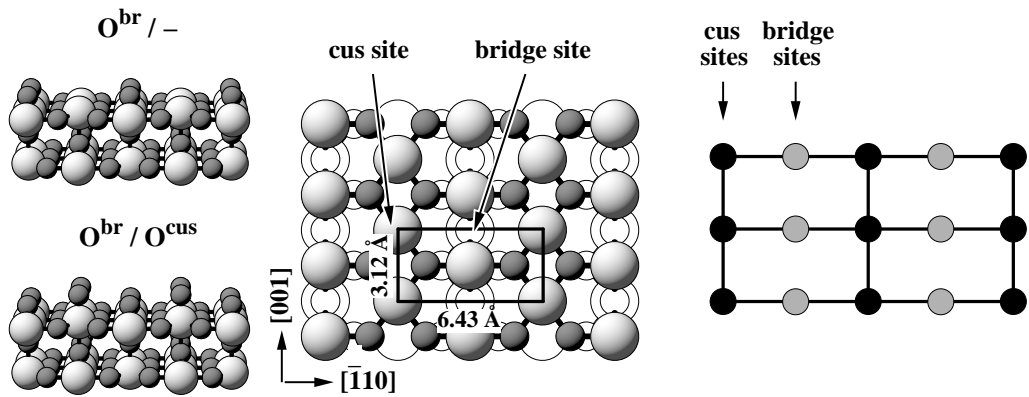}}
\caption{\label{ruo2_lattice}
Top view of the RuO${}_2$(110) surface showing the two prominent adsorption sites (bridge and cus), which continue the bulk-stacking sequence (central panel). When focusing on these two site types, the surface can be coarse-grained to the lattice model shown schematically to the right. Additionally shown are two perspective views to the left, exemplifying what the atomic structure of the surface looks like, if all bridge sites are occupied with oxygen atoms and the cus sites remain empty (O$^{\rm br}$/- termination, top left panel), or if oxygen atoms occupy both site types (O$^{\rm br}$/O$^{\rm cus}$ termination, bottom left panel). Ru = light, large spheres, O = dark, medium spheres. Atoms lying in deeper layers have been whitened in the topview for clarity.}
\end{figure*}

Diffusion processes of adsorbed particles on the lattice can also be treated within a TST framework. The rate for such hops from one site (of type $st$) to another site (of type $st^\prime$) is then given by
\begin{equation}
r^{\rm diff}_{st,st^\prime,i}(T) \;=\;
f^{\rm diff,TST}_{st,st^\prime,i}(T) \; \left(\frac{k_BT}{h}\right)
\; {\rm exp} \left(- \frac{\Delta E^{\rm diff}_{st,st^\prime,i}}{k_B T}\right) \quad ,
\label{diffrate}
\end{equation}
where
\begin{equation}
f^{\rm diff,TST}_{st,st^\prime,i}(T) \;=\; \frac{q^{\rm vib}_{{\rm TS}(st,st^\prime,i)}}{q^{\rm vib}_{st,i}} \quad .
\label{f_difftst}
\end{equation}
$\Delta E^{\rm diff}_{st,st^\prime,i}$ denotes the maximum barrier along the MEP between the two sites, $q^{\rm vib}_{{\rm TS}(st,st^\prime,i)}$ the corresponding partition function at this barrier, and $q^{\rm vib}_{st,i}$ the partition function at the bound state in site $st$ as before.

Similar to the process types covered before, fulfillment of microscopic reversibility is also here a crucial issue. This is trivially given for diffusion between identical site types, since $f^{\rm diff,TST}_{st,st,i}(T)$ and $\Delta E^{\rm diff}_{st,st,i}$ are identical for the forward and backward jump. For diffusion between different site types, e.g. bridge to cus site and the reversed process cus to bridge site in the RuO$_2$(110) example, this is no longer the case. If approximations are employed here, they have to be made consistently for both the forward and backward diffusion process.

\subsection{First-principles parameters for the CO oxidation over RuO$_2$(110)}

\subsubsection{Lattice model}

The rutile RuO${}_2$(110) surface exhibits a rectangular $(1 \times 1)$ unit-cell as shown in Fig. \ref{ruo2_lattice}. In this orientation, the bulk stacking sequence can be expressed as a series of O-(Ru$_2$O$_2$)-O trilayers. \cite{reuter02a,reuter03a} Depending at which layer the sequence is truncated, three different $(1 \times 1)$ surface terminations result, and the middle panel of Fig. \ref{ruo2_lattice} shows a topview of the surface atomic structure when the cut yields a top (Ru$_2$O$_2$) layer. All oxygen atoms in this topmost (Ru$_2$O$_2$) layer maintain their threefold bulk-like coordination to Ru atoms and are thus tightly bound.\cite{reuter03a} Binding to the undercoordinated surface Ru atoms, on the other hand, gives rise to two prominent adsorption sites per surface unit-cell, namely a bridge (br) site bridging two surface Ru atoms and a so-called coordinatively unsaturated (cus) site atop of one surface Ru atom, cf. Fig. \ref{ruo2_lattice}. The rutile bulk-stacking sequence would be continued by oxygen atoms occupying first all bridge sites (leading to the so-called O${}^{\rm br}$/- termination, upper left panel in Fig. \ref{ruo2_lattice}) and then all cus sites (O${}^{\rm br}$/O${}^{\rm cus}$ termination, lower left panel), before another (Ru$_2$O$_2$) plane would start a new sequence. The O${}^{\rm br}$/- termination is routinely observed experimentally after high temperature anneals in ultra-high-vacuum (UHV), whereas the O${}^{\rm br}$/O${}^{\rm cus}$ termination is most stable a high oxygen pressures.\cite{reuter02a,reuter03a,over03} Extensive theoretical work over the last years has shown that O and CO adsorption at other sites of the surface is energetically significantly less favorable than adsorption  at bridge and cus sites \cite{reuter02a,reuter03a,over03}, which is consistent with all presently available experimental data \cite{over03,boettcher99,kim01a,wang01,kim01b,seitsonen01,kim03,paulus03}.

In view of these findings we choose the following model of the RuO$_2$(110) system as basis of our kMC simulations: The surface is described by a periodic lattice with a rectangular surface unit-cell, each cell containing one bridge and one cus site. The arrangement of these sites is such that along the $[\bar{1}10]$ direction, i.e. the long side of the rectangle, bridge and cus sites are alternating, while along the short direction of the rectangle (representing $[001]$) only like sites align. This leads to a characteristic stripe pattern with alternating rows of bridge and cus sites as schematized in the right panel of Fig. \ref{ruo2_lattice}. Each of the sites in the lattice can be either empty, or be occupied by one O or CO. Since this comprises the possibility of bridge sites not being occupied with oxygen atoms, we note that our treatment includes thus implicitly the effect of the frequently discussed O surface vacancies in the UHV O${}^{\rm br}$/- termination.

In the context of catalytic CO oxidation, the kMC simulations need to consider adsorption, desorption, diffusion and chemical reactions on this lattice. CO adsorption into vacant cus or bridge sites is non-dissociative, while oxygen adsorption is dissociative and requires two vacant neighboring sites, i.e. a br-br, cus-cus, or br-cus pair, cf. Fig. \ref{ruo2_lattice}. Together with the corresponding (time-reversed) desorption processes this leads already to 10 different process types. Diffusion of adsorbed O or CO is modeled as hops to nearest neighbor lattice sites, and since adsorbate diffusion can go br-to-br, br-to-cus, cus-to-cus, or cus-to-br, this adds another 8 processes. Product CO$_2$ molecules may finally be formed by reaction of adsorbed O and CO in neighboring sites. There are therefore four different reaction processes possible, namely O$^{\rm br}$+CO$^{\rm br}$, O$^{\rm br}$+CO$^{\rm cus}$, O$^{\rm cus}$+CO$^{\rm br}$ and O$^{\rm cus}$+CO$^{\rm cus}$. At the temperatures of interest to our study the formed CO$_2$ desorbs immediately \cite{wang02b}, so that one reaction event is modeled as associative CO$_2$ desorption and leads to two vacant sites on the lattice. In order to fulfill detailed balance, also the back reaction, i.e. dissociative CO$_2$ adsorption, has then to be included, adding another four processes to the list. In total, our kMC simulations on this lattice model consider therefore 26 different elementary processes, for each of which DFT calculations were carried out to provide the parameters needed to determine the process rates via the equations derived in Section IIB.

\subsubsection{DFT setup}

The DFT calculations use the full-potential linear augmented plane wave approach (FP-LAPW) \cite{blaha99,kohler96,petersen00} together with the generalized gradient approximation (GGA) for the exchange-correlation functional \cite{perdew96}. The reference calculations for the free gas phase molecules, as well as all calculations addressing O$_2$ adsorption and desorption barriers were done in a spin-polarized mode. Extensive test calculations confirmed that in all adsorbed configurations, including surface diffusion, the spin is quenched at the metallic RuO$_2$(110) surface, and the corresponding computations were carried out in a non-spin-polarized way. The RuO${}_2$(110) surface is modeled in a supercell geometry, employing a symmetric slab consisting of three rutile O-(Ru$_2$O$_2$)-O trilayers. All atomic positions within the outermost trilayer were fully relaxed and a vacuum region of $\approx$ 11 {\AA} ensures the decoupling of the surfaces of consecutive slabs.

The FP-LAPW basis set parameters are: $R_{\rm{MT}}^{\rm{Ru}}=$1.8 bohr, $R_{\rm{MT}}^{\rm{O}}=$1.1 bohr, $R_{\rm{MT}}^{\rm{C}}=$1.0 bohr, wave function expansion inside the muffin tins up to $l_{\rm{max}}^{\rm{wf}} = 12$, potential expansion up to $l_{\rm{max}}^{\rm{pot}} = 4$. For the RuO${}_2$(110) slabs the 
$(1 \times 1)$ Brillouin zone (BZ) integration was performed using a $(5 \times 10 \times 1)$ Monkhorst-Pack grid with 50 (15) {\bf k}-points in the full (irreducible) part of the BZ. To obtain the same sampling of the reciprocal space for bigger surface cells, this number is reduced accordingly. The energy cutoff for the plane wave representation in the interstitial region between the muffin tin spheres was $E^{\rm max}_{\rm wf}$ = 20 Ry for the wave functions and $E^{\rm max}_{\rm pot}$ = 169 Ry for the potential. 

The computational setup is thus exactly the same as the one employed in our previous work on RuO$_2$(110).\cite{reuter02a,reuter03a} Its numerical accuracy has already been detailed there, and with respect to the present study it is characterized by a $\pm 150$\,meV uncertainty in the binding energies and activation barriers, and a $\pm 50$\,meV uncertainty in relative binding energy differences for the same species (e.g. at different coverages). Although already this setup led to a massive computational demand, the absolute numbers are therefore not well converged. We confirmed, however, that the correct energetic ordering is obtained \cite{kiejna05}, which is what largely determines the results discussed in this work. The effect of the uncertainty in the kMC rates due to the errors in the energetics will be further discussed in section IIIC. This then also comprises the additional uncertainty introduced by the more basic deficiency of density-functional theory, namely the approximate nature of the employed exchange-correlation functional, here the GGA.

\subsubsection{Adsorption, Desorption and Reaction}

Within this setup we first address processes, which change the number of oxygen atoms or CO molecules adsorbed at lattice sites. This naturally includes adsorption and desorption of the reactants, but as mentioned above also reaction processes leading to CO$_2$ formation, which are treated as associative CO$_2$ desorption. The energetics along the MEP summarized in Fig. \ref{mep} play a decisive role in determining the rates of these processes. In general, all corresponding quantities appearing in eq. (\ref{mep_relation}) depend not only on the site types actually involved in the process, but also on the local environment around them, i.e. on the occupation of nearby lattice sites. Formally such adsorbate-adsorbate interactions can be taken into account through a lattice gas Hamiltonian (LGH), expanding the total energy in terms of on-site energies and lateral interactions of pair or higher many-body type. \cite{reuter05a}

\begin{table}

\caption{\label{tab1}
Binding energies, $E_b$ (in eV), in $(1 \times 1)$ surface configurations involving only one species (O or CO) and occupation of all sites of one site type (br or cus), i.e. a coverage of 1 monolayer (ML). Using surface unit-cells of varying size, this total coverage is varied between 1/3 and 1 monolayer (ML), and the resulting variations in binding energy with respect to the value at 1\,ML are given.}
\begin{ruledtabular}
\begin{tabular}{c|c|cc|c}
 &                &\multicolumn{2}{c|}{$\Delta E_b$ due to} & $\Delta E_b$ due to    \\
 &   $E_b$        &\multicolumn{2}{c|}{interactions}    & interactions    \\
 &                &\multicolumn{2}{c|}{along rows}       & across rows     \\ \hline
 & $(1 \times 1)$ & $(1 \times 2)$ & $(1 \times 3)$ & $(2 \times 1)$  \\ 
 & 1\,ML          & $\frac{1}{2}$\,ML & $\frac{1}{3}$\,ML & $\frac{1}{2}$\,ML \\
\hline
O$^{\rm br}$  &$-2.44$ & +0.11            & +0.09            & +0.01       \\
O$^{\rm cus}$ &$-1.08$ &$-0.06$           &$-0.04$           &$-0.06$      \\
CO$^{\rm br}$ &$-1.64$ &$-0.02$           &$-0.02$           & +0.08       \\
CO$^{\rm cus}$&$-1.31$ &$-0.07$           &$-0.07$           & +0.07         \\
\end{tabular}
\end{ruledtabular}
\end{table}

In order to assess the variations in the process energetics that arise due to lateral interactions we concentrate first on the binding energy, i.e. the minimum of the MEP. In the sense of a LGH, lateral interactions affecting this quantity may be classified into interactions between like (i.e. br-br, cus-cus) and unlike (i.e. br-cus) sites, as well as into interactions between like (i.e. O-O, CO-CO) and unlike (i.e. O-CO) species. Starting with the interactions between like sites and like species, we computed a series of surface configurations involving always only one surface species and occupation of only one site type at various concentrations. Specifically, we employed $(1 \times 1)$, $(1 \times 2)$ and $(1 \times 3)$ surface unit-cells to describe atomic arrangements, in which every, every second or only every third site along the rows in the $[001]$ direction, cf. Fig. \ref{ruo2_lattice}, is occupied. Calculations in $(2 \times 1)$ cells allowed accordingly to change the concentration in the $[\bar{1}10]$ direction across the rows. The resulting binding energies are summarized in Table \ref{tab1}. They show only minor variations, indicating overall rather weak lateral interactions of O$^{\rm br}$-O$^{\rm br}$, O$^{\rm cus}$-O$^{\rm cus}$, CO$^{\rm br}$-CO$^{\rm br}$, and CO$^{\rm cus}$-CO$^{\rm cus}$ type.

We interpret this finding as a consequence of the rather open crystal structure of the RuO$_2$(110) surface. For a bridge site, already a nearest neighbor bridge site across the rows and a second nearest neighbor bridge site along the rows is more than 6\,{\AA} away, cf. Fig. \ref{ruo2_lattice}, and neither of these sites involves a coordination to surface atoms already involved in the direct bonding to the given bridge site. For a cus site and its neighboring cus sites, this situation is exactly the same. A sharing of surface atoms at close distance occurs only upon occupation of the nearest neighbor bridge site along the bridge rows, explaining the slightly larger variation of the binding energy for a surface atomic arrangement in which oxygen atoms occupy such neighboring sites, cf. Table \ref{tab1}. It is interesting to note that the corresponding binding energy variation is masked for the case of adsorbed CO$^{\rm br}$ molecules, since there the bonding geometry changes between a symmetric and an asymmetric configuration, depending on whether neighboring bridge sites are occupied or not. This has been discussed extensively in Refs. \onlinecite{seitsonen01} and \onlinecite{paulus03}, and our DFT results concerning the structural and energetic details are in full agreement with the data reported there.

\begin{table}
\caption{\label{tab2}
Binding energies, $E_b$ (in eV), of O and CO surface species adsorbed in a $(1 \times 1)$ configuration and with varying occupations at the other site type within the surface unit cell.}
\begin{ruledtabular}
\begin{tabular}{c|ccc}
              & \multicolumn{3}{c}{\hspace{-1cm}occupation at nearest neighbor bridge site}\\
                &  empty  &  O$^{\rm br}$ &  CO$^{\rm br}$   \\ \hline

O$^{\rm cus}$   &  -1.08  &    -0.99      &    -1.01         \\
CO$^{\rm cus}$  &  -1.31  &    -1.26      &    -1.26         \\ \hline \hline
                & \multicolumn{3}{c}{\hspace{-1cm}occupation at nearest neighbor cus site}\\
                &  empty  &  O$^{\rm cus}$&  CO$^{\rm cus}$  \\ \hline
O$^{\rm br}$    &  -2.44  &    -2.23      &    -2.37         \\
CO$^{\rm br}$   &  -1.64  &    -1.57      &    -1.58         \\
\end{tabular}
\end{ruledtabular}
\end{table}

The variations in the binding energies due to lateral interactions between like species and like sites are always well within $\pm 150$\,meV compared to the value in a $(1 \times 1)$ arrangement. Similar findings were obtained in calculations for atomic arrangements involving again only occupation of one site type (either bridge or cus), but having O and CO simultaneously present at the surface. Again, the binding energy variations were well within $\pm 150$\,meV, indicating that also interactions of the type O$^{\rm br}$-CO$^{\rm br}$ and O$^{\rm cus}$-CO$^{\rm cus}$ are rather small. This leaves only interactions between bridge and cus sites, both between like and unlike species. Calculations with sparse atomic arrangements in $(1 \times 2)$ and $(1 \times 3)$ surface unit-cells showed in this case that br-cus interactions between second and third nearest neighbor sites at distances larger than 4.4\,{\AA} are negligibly small. The binding energy variations become only somewhat more appreciable, when nearest neighboring bridge and cus sites are involved. This is illustrated in Table \ref{tab2} for arrangements with $(1 \times 1)$ periodicity, showing that primarily adsorbed oxygen atoms exhibit some variation in the bond strength, depending on whether the other site type within the surface unit-cell is either empty or occupied by O or CO.

In summary, our calculations reveal that lateral interactions, affecting the binding of O and CO to the RuO$_2$(110) surface, are small. The only significant interactions are with nearest neighbor lattice sites at an around 3\,{\AA} distance, cf. Fig. \ref{ruo2_lattice}. Along the rows in the $[001]$ direction this is a like site (i.e. br-br or cus-cus), whereas across the rows in the $[\bar{1}10]$ direction this is the interaction with the next unlike site (i.e. br-cus). We expect such interactions to play a role in stabilizing ordered adsorbate superstructures at the surface at low temperatures. With $E_{b}({\rm O}^{\rm cus}) = -1.0\pm 0.15$\,eV, $E_{b}({\rm O}^{\rm br}) = -2.3\pm 0.15$\,eV, $E_{b}({\rm CO}^{\rm cus}) = -1.3\pm 0.15$\,eV, and $E_{b}({\rm CO}^{\rm br}) = -1.6\pm 0.15$\,eV in all calculated test arrangements, we recognize, however, that the variations are still rather small compared to the other energies in this system. For the present study, we therefore neglect adsorbate-adsorbate interactions and employ the just listed four $E_b$ values to describe the binding of O and CO at the two sites throughout. 

\begin{figure}[t!]
\scalebox{0.35}{\includegraphics{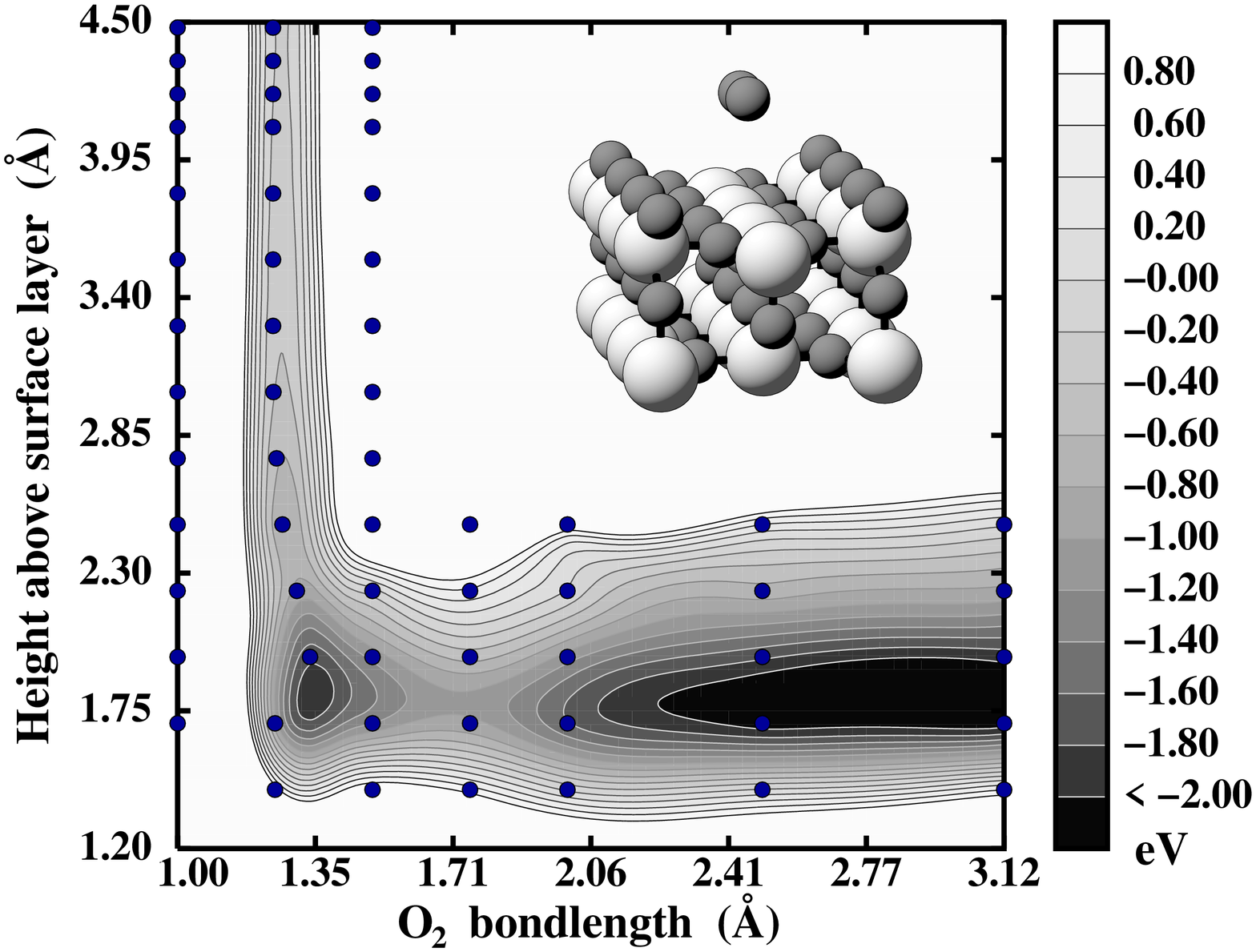}}
\caption{Potential-energy surface (PES) for associative oxygen desorption from two neighboring cus sites into an O$_2$ molecule with orientation parallel to the surface and aligned along the [001] trench (as schematically shown in the inset). The height above the topmost (Ru$_2$O$_2$) surface and the distance between the two O atoms are used as coordinates of the plot. For every (height, bondlength) of the parallel O$_2$ molecule, all atomic positions of the oxide substrate are fully relaxed. Actually calculated points are indicated, and the energy zero corresponds to a free O$_2$ molecule far above the surface.}
\label{pes_elbow}
\end{figure}

Considering the results obtained for the binding energies, we do not expect the other parts of the MEP to be much more affected by lateral interactions. All calculations of energy barriers along adsorption, desorption and reaction pathways were correspondingly carried out using $(1 \times 2)$ surface unit-cells, this way ensuring that at the relevant transition state geometries the distances to all periodic images are always larger than 6\,{\AA}. 
Starting the description of such barrier calculations with the adsorption and desorption of the reactants, we first address the associative oxygen desorption from two neighboring cus sites. For this we computed the PES, when lifting the associating molecule from the O$^{\rm br}$/O$^{\rm cus}$ termination, i.e. using the equal height of both O$^{\rm cus}$ atoms and the distance betweem them as reaction coordinates, while minimizing the energy at each point with respect to all remaining degrees of freedom. As apparent from Fig. \ref{pes_elbow}, there is a shallow molecular chemisorption well at a height of 1.8\,{\AA}, which has already been described in previous DFT calculations \cite{kim01a}. Above this well the energy rises, however, smoothly to the level of a free gas phase molecule, reflecting that for this molecular orientation dissociation is non-activated. 

Equivalent findings were also obtained when lifting a pair of O atoms from neighboring bridge-bridge sites and from neighboring bridge-cus sites, as well as when lifting up an upright CO molecule from a bridge and a cus site, i.e. none of these cases showed an activation barrier. While these calculations for fixed molecular orientation and lateral position over the unit-cell do, of course, not account for the variations in the full high-dimensional PES for adsorption and desorption, they still show that the maximum barrier height along the MEP connecting the gas phase and adsorbed state must be zero. As discussed in section IIB.1 we correspondingly use
\begin{eqnarray}
\nonumber
\Delta E^{\rm ad}_{\rm O, cus-cus} &=& \Delta E^{\rm ad}_{\rm O, br-br} \;=\;
\Delta E^{\rm ad}_{\rm O, cus-br} \;= \\
\Delta E^{\rm ad}_{\rm CO, cus} &=& \Delta E^{\rm ad}_{\rm CO, br} \;=\; 0
\end{eqnarray}
in the determination of the local sticking coefficient through eq. (\ref{stick_general}). This leaves then only the quantities $f^{\rm ad}_{st,i}(T)$ and $A_{st,i}$ to entirely determine the rates of all 5 possible adsorption processes of reactants, which in turn also determines the corresponding desorption rates through eq. (\ref{equilibrium2}), using $q^{\rm vib}_{st,i} = 1$.

The quantity $f^{\rm ad}_{st,i}(T)$ accounts for a reduction of the sticking probability, if particles with certain initial states (lateral positions over the unit-cell, internal degrees of freedom) are not steered along the MEP and are then potentially reflected by a higher barrier. In the temperature range of interest to the present study ($\sim 300 - 600$\,K), impinging thermal molecules are still comparably slow, and we expect them to be efficiently steered along non-activated pathways in the high-dimensional PES. As described above, an accurate determination of the therefore expected value of $f^{\rm ad}_{st,i}(T) \approx 1$ would require computationally highly demanding dynamical simulations of a statistically relevant number of trajectories for the impinging molecules. \cite{gross98} As we will discuss in section IIIC, uncertainties in the individual rates by a factor of 10 to 100 do not affect the reported conclusions. Since $f^{\rm ad}_{st,i}(T)$ enters the rates only linearly, we therefore simply set it to unity, and with the same argument, partition the active areas for adsorption equally over the various sites. Since there are four ways for an O$_2$ to dissociate into site pairs in a $(1 \times 2)$ unit-cell, this leads to $A_{\rm O, cus-cus} = A_{\rm O, br-br} = A_{\rm O, cus-br} = 1/2 A_{uc}$, and with the same argument we obtain $A_{\rm CO, cus} = A_{\rm CO, br} = 1/2 A_{uc}$ for CO adsorption. With this choice of prefactors the sum of local sticking coefficients over all site types in the surface unit-cell has the correct upper bound, and through the use of eq. (\ref{equilibrium2}) to determine the corresponding desorption rates, detailed balance is also ensured.

Having determined the rates of the 10 possible adsorption and desorption processes of the reactants, we are left with surface reactions as the final event type changing the number of O or CO adsorbed at lattice sites. Similar to the procedure employed for the desorption events we evaluate the energetics by mapping out the PES along suitable reaction coordinates, fully relaxing all remaining degrees of freedom. For two of the four possible reaction processes, this was already described in detail in a preceding publication \cite{reuter03a}, and exactly the same methodology was now used to locate the remaining two transition states. The resulting reaction barriers (or equivalently associative CO$_2$ desorption barriers) are: $\Delta E^{\rm des}_{\rm O^{\rm br} + CO^{\rm br}} = 1.54$\,eV, $\Delta E^{\rm des}_{\rm O^{\rm br} + CO^{\rm cus}} = 1.25$\,eV, $\Delta E^{\rm des}_{\rm O^{\rm cus} + CO^{\rm cus}} = 0.89$\,eV, and $\Delta E^{\rm des}_{\rm O^{\rm cus} + CO^{\rm br}} = 0.76$\,eV. Since in this case all four reaction processes have clear defined transition states, it is useful to resort to eq. (\ref{desrate}) to fix the process rates. For this, we employ $f^{\rm des,TST}_{st,i} = 0.5$, which includes a value of $A_{st,i} = 1/2 A_{uc}$, again simply equally partitioning the active areas for the corresponding time-reversed dissociative CO$_2$ adsorption processes. Effectively, we therefore set $q^{\rm vib}_{{\rm TS}(st,i)} = q^{\rm vib}_{st,i}$, cf. eq. (\ref{f_destst}), and a more rigorous approach would be to explicitly evaluate the computationally more demanding partition function $q^{\rm vib}_{{\rm TS}(st,i)}$ at the barrier. Test calculations in the harmonic approximation indicate that the uncertainty introduced by the employed approximation translates into variations of the desorption rate by a factor of 10 (at most 100), i.e. an uncertainty that is of the same order as the one in the determination of the reactant adsorption and desorption process rates. 

Together with the reactant binding energies, the desorption barriers can be used to obtain the activation barriers for the corresponding dissociative CO$_2$ adsorption processes. This yields in all four cases sizable barriers, which when input into eq. (\ref{stick_general}) lead to vanishingly small values below $10^{-12}$ for the local sticking coefficient at all temperatures of interest. Regardless of the exact CO$_2$ partial pressures building up over the operating catalyst surface, dissociative CO$_2$ readsorption into vacant sites at the surface will therefore be completely negligible compared to the non-activated adsorption of oxygen and CO.

\subsubsection{Diffusion}

\begin{table}
\caption{\label{tab3}
Binding energies, $E_b$, for CO and O (with respect to (1/2)O$_2$) 
at bridge and cus sites, cf. Fig. \ref{ruo2_lattice}, and diffusion energy
barriers, $\Delta E^{\rm diff}$, to neighboring bridge and cus sites, as used in the kMC simulations. The desorption barriers are given for unimolecular and for associative desorption with either O$^{\rm cus}$ or O$^{\rm br}$. This includes therefore surface reactions forming CO$_2$, which are considered as associative desorption of an adsorbed O and CO pair. All values are in eV.}
\begin{ruledtabular}
\begin{tabular}{l|c@{\quad}|ccc@{\quad}|cc}

  &$E_b$ & \multicolumn{3}{c|}{$\Delta E^{\rm des}$}  &
           \multicolumn{2}{c}{$\Delta E^{\rm diff}$} \\
  &      & unimol. & with O$^{\rm br}$ & with O$^{\rm cus}$ & to br & to cus \\ \hline
CO$^{\rm br}$  & -1.6 & 1.6 & 1.5 & 0.8     & 0.6 & 1.6 \\
CO$^{\rm cus}$ & -1.3 & 1.3 & 1.2 & 0.9     & 1.3 & 1.7 \\
O$^{\rm br}$   & -2.3 & $-$ & 4.6 & 3.3     & 0.7 & 2.3 \\
O$^{\rm cus}$  & -1.0 & $-$ & 3.3 & 2.0     & 1.0 & 1.6 \\ 
\end{tabular}
\end{ruledtabular}
\end{table}

With 18 process rates determined, we now turn to the 8 diffusion processes. In line with the computations detailed above, the corresponding barriers were obtained using $(1 \times 2)$ unit-cells and mapping the PES along the high-symmetry line connecting the initial and final site of the diffusion process, while fully relaxing all remaining degrees of freedom. Test calculations checking on the effect of lateral interactions with the periodic images were performed in suitably extended surface unit-cells, namely $(1 \times 3)$ unit-cells for diffusion processes along the trenches (i.e. cus-to-cus and br-to-br), and $(2 \times 2)$ for diffusion across the trenches (i.e. cus-to-br and br-to-cus). The observed $< \pm 80$\,meV variations with respect to the barriers obtained in the standard $(1 \times 2)$ cells reflect just as in the case of the binding energies that lateral interactions with atoms beyond 6\,{\AA} distance are insignificant at this surface. Since the binding energy calculations indicated slightly larger lateral interactions between nearest-neighbor bridge and cus sites, all diffusion processes along the trenches were furthermore calculated with different occupations of the corresponding other sites, e.g. the oxygen cus-to-cus barrier was computed with O$^{\rm br}$ or CO$^{\rm br}$ present at the neighboring bridge sites, or with the bridge sites empty. Again, the barrier variations were with $\pm 130$\,meV only modest. Therefore neglecting adsorbate-adsorbate interactions, we employed the energetic barriers listed in Table \ref{tab3} throughout to describe the eight possible types of diffusion events on the lattice. For completeness Table \ref{tab3} also compiles the energetic barriers for all other processes as used in the kMC simulations.

To fully fix the diffusion process rates in eq. (\ref{diffrate}), we use $f^{\rm diff, TST}_{st,st^\prime} = 1$, i.e. like in the surface reaction case we again approximate $q^{\rm vib}_{{\rm TS}(st,st^\prime,i)} = q^{\rm vib}_{st,i}$. From test calculations in the harmonic TST approximation we estimate that this has an uncertainty of a factor of 10. We will see below that the ensuing (already quite low) uncertainty in the corresponding rates is for this process class particularly inconsequential. Due to the high barriers, most surface diffusion processes have such low rates that they occur only very rarely anyway, and under the catalytically most relevant high-pressure conditions virtually all surface sites are occupied, thereby additionally blocking the remaining diffusion events.

\subsection{kMC setup for steady-state catalysis}

Using the now completely determined 26 process rates, kinetic Monte Carlo runs were performed for $(T,p_{\rm CO},p_{\rm O_2})$ conditions covering the temperature range from $T = 300$\,K to $T=800$\,K and partial pressures from $10^{-15}$ to $10^5$ atmospheres. In each simulation for fixed $(T,p_{\rm CO},p_{\rm O_2})$, the evolution of the system from an arbitrary initial lattice configuration is first calculated until steady-state conditions are reached, and the targeted average surface quantities are thereafter evaluated over sufficiently long time spans to reach a 1\,\% statistical precision. The simulations were routinely repeated using other initial lattice configurations to verify that the true dynamic steady-state was reached. Besides the average surface populations, this kind of statistical analysis yields e.g. the average frequencies of the various elementary processes, and through the average frequency of all surface reaction events then also the absolute TOF under the chosen environmental conditions.

\begin{figure}
\scalebox{0.65}{\includegraphics{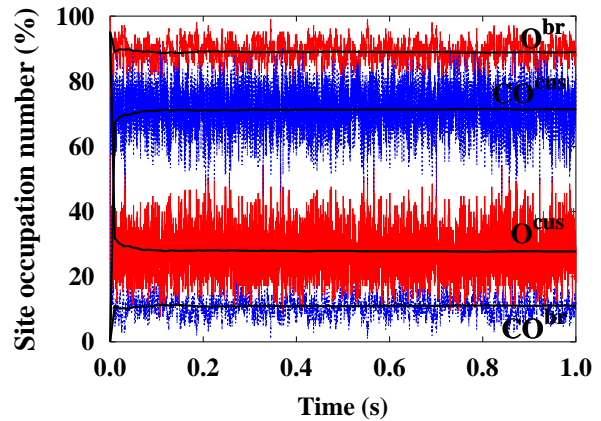}}
\caption{\label{evolution}
(Color online) Time evolution of the site occupation by O and CO of the two prominent adsorption sites, bridge and cus, cf. Fig. \ref{ruo2_lattice}. The temperature and pressure conditions chosen ($T=600$\,K, $p_{\rm CO} = 20$\,atm, $p_{\rm O_2} = 1$\,atm) correspond to optimum catalytic performance. Under these conditions kinetics builds up a steady-state surface population in which O and CO compete for either site type at the surface, as reflected by the strong fluctuations in the site occupations. Note the extended time scale, also for the ``induction period'', until the steady-state populations are reached when starting from a purely oxygen covered surface.}
\end{figure}

The procedure is illustrated in Fig. \ref{evolution}, where the actual and average surface populations of O$^{\rm br}$, O$^{\rm cus}$, CO$^{\rm br}$ and CO$^{\rm cus}$ are plotted, when starting from a fully oxygen covered surface (O$^{\rm br}$/O$^{\rm cus}$ termination). Despite notable fluctuations, which we will discuss in more detail below, well defined average values for all surface species are finally obtained. We note that although the individual elementary process dynamics takes place on a picosecond time scale, the ``induction period'' until the steady-state populations are reached is of the order of a tenth of a second. At the lower temperatures of interest to our study, namely 350\,K, this becomes even more pronounced. Due to the decelerated process rates, the real times spanned in the simulations (at an equivalent number of kMC steps) can then be orders of magnitude longer. Only the efficient time coarse-graining underlying kMC algorithms makes it possible to reach such time scales, while still accounting for the full atomic-scale correlations, fluctuations and spatial distributions.

\begin{figure}
\scalebox{0.49}{\includegraphics{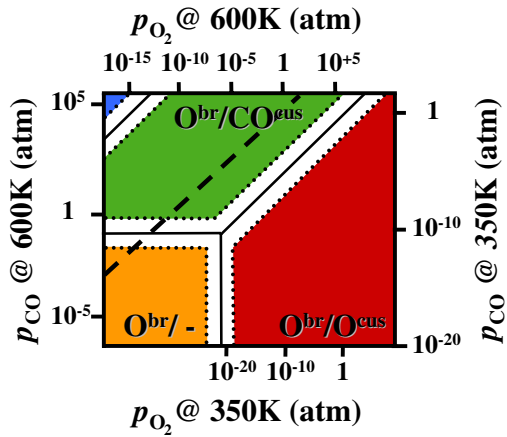}}
\caption{\label{surfpop_thermo}
(Color online) Steady-state surface structures obtained by first-principles kMC simulations, where the surface reaction events are switched off (see text). In all non-white areas, the average site occupation is dominated ($> 90$\,\%) by one species, i.e. either O, CO or empty sites ($-$). The labels correspondingly characterize the surface populations by indicating this majority species at the bridge and cus sites. The unlabeled small region just appearing in the upper left edge of the figure corresponds to a fully CO-covered surface, CO$^{\rm br}/{\rm CO}^{\rm cus}$. Note, that in the thermodynamic case the stability regions of the various surface phases depend only on the gas phase chemical potentials $\mu_i$ ($i =$\,O,CO), leading to a simple scaling of the pressure axes at different temperatures, cf. eq. (\ref{mu_gas}). The only exception to this is the width of the white coexistence regions, which increases with temperature. The one shown here corresponds to 600\,K. Above the dashed line bulk RuO$_2$ is thermodynamically unstable against CO-induced decomposition (see text).}
\end{figure}

The simulations were carried out on a system with $(20 \times 20)$ surface sites (200 bridge and 200 cus sites) and periodic boundary conditions. Test runs for a $(30 \times 30)$ system gave identical results, and even then the computational effort of the kMC simulations themselves was still insignificant compared to the first-principles calculations required to evaluate the process rates. Before commencing with the real simulations, a useful check of the detailed balance of our setup was furthermore to run simulations with all surface reaction events switched off. The DFT-kMC approach is then equivalent to the computationally much less demanding ``constrained equilibrium'' approach of atomistic thermodynamics \cite{reuter03a,reuter05a} and we indeed reproduce exactly our earlier results, which were based on a sub-set of the present DFT data \cite{reuter02a,reuter03a}. The resulting average surface populations are shown in Fig. \ref{surfpop_thermo} to facilitate direct comparison with the kinetic data and are discussed in more detail below.

\section{Results}

\subsection{Steady-state surface composition and stoichiometry}

\begin{figure*}
\scalebox{0.30}{\includegraphics{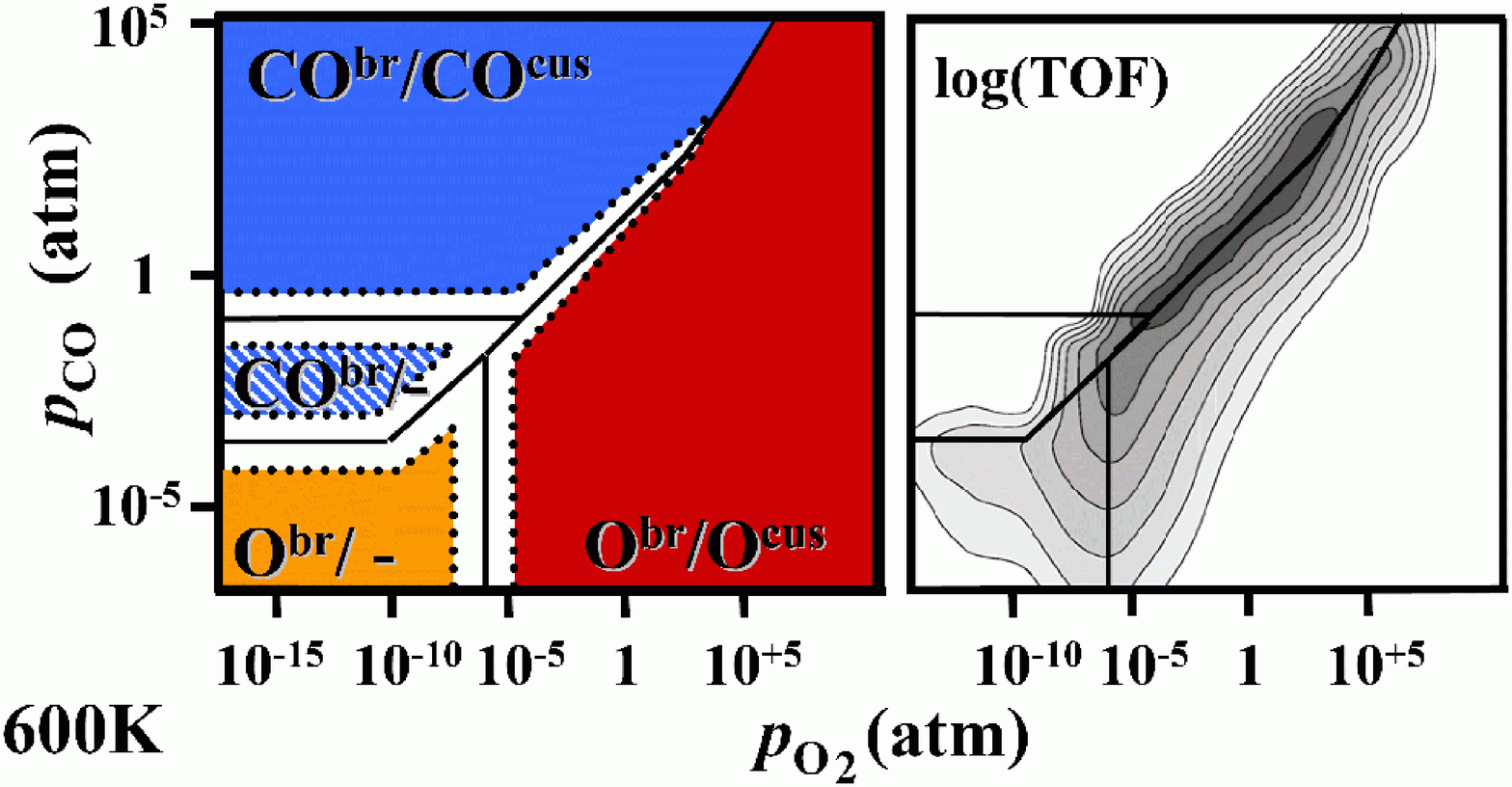}}
\caption{\label{TOF_600}
(Color online) Left panel: Steady-state surface structures obtained by first-principles kMC simulations at $T = 600$\,K (see Fig. \ref{surfpop_thermo} for an explanation of the labeling). Right panel: Map of the corresponding turn-over frequencies (TOFs) in ${\rm site}^{-1} {\rm s}^{-1}$\; ($10^{15}\; {\rm cm}^{-2}{\rm s}^{-1}$): White areas have a TOF\,$< 10^{-4}\; {\rm site}^{-1} {\rm s}^{-1} \;\;(< 10^{11}\; {\rm cm}^{-2}{\rm s}^{-1})$, and each increasing gray level represents one order of magnitude higher activity. Thus, the darkest gray region corresponds to TOFs above $10^{2}\; {\rm mol.}\,{\rm site}^{-1} {\rm s}^{-1}\;\; (10^{17}\; {\rm cm}^{-2}{\rm s}^{-1})$, while in a narrow $(p_{\rm CO},p_{\rm O_2})$ region the TOFs actually peak over $10^{3}\; {\rm site}^{-1} {\rm s}^{-1}\;\; (10^{18}\; {\rm cm}^{-2}{\rm s}^{-1})$. The plots are based on $\approx 400$ kMC simulations for different $(p_{\rm CO},p_{\rm O_2})$ conditions.}
\end{figure*}

We start by using our first-principles kMC simulations to obtain the steady-state average surface populations in a wide range of $(T,p_{\rm CO},p_{\rm O_2})$ conditions that are of interest in the catalytic context. For this system this is specifically the temperature window between $\sim 350 - 600$\,K, where high catalytic activity has been reported experimentally for varying pressures ranging from UHV up to technologically relevant values of the order of atmospheres \cite{wang02,over03,peden86}. Figure \ref{TOF_600} summarizes the results for $T = 600$\,K and varying O$_2$ and CO partial pressures. Under most pressure conditions, the site occupation is found to be dominated by more than 90\,\% by one species (O or CO) or vacant sites ($-$). This gives rise to four defined steady-state surface structures. For the lowest O$_2$ and CO partial pressures this is the stoichiometric O$^{\rm br}/-$ termination, where virtually all bridge sites are occupied by oxygen atoms and all cus sites are empty. Higher O$_2$ partial pressures stabilize the O-rich O$^{\rm br}/{\rm O}^{\rm cus}$ termination, where now also all cus sites are occupied by O atoms. Both terminations are routinely observed experimentally under UHV conditions, the prior directly after high-temperature annealing and the latter after oxygen post-exposure.\cite{over03,boettcher99,kim01a} At increased CO partial pressures, we find two additional kinds of structures, CO$^{\rm br}/-$ and CO$^{\rm br}/{\rm CO}^{\rm cus}$, both of which have recently also been characterized in UHV following different preparation recipes \cite{seitsonen01,paulus03,kim03}. 

The transition from one structure to the other with changing pressures is not abrupt, but occurs over a finite pressure range. Under such conditions, more than one species (and/or vacancies) occupies a significant fraction ($> 10$\,\%) of either surface site type, and we delimit such coexistence regions in Fig. \ref{TOF_600} by dotted lines. If the system were in equilibrium with the surrounding gas phase, such coexistence regions between stable surface phases could be described by evaluating the configurational entropy, and in the present case due to the absence of lateral interactions this would simply be Langmuir adsorption isotherms \cite{masel96}. In fact, when switching off the reaction events in the kMC simulations the simulated average occupations in coexistence regions are indeed identical to those resulting from the appropriate analytical Langmuir expression, since the kMC algorithm automatically accounts for the configurational entropy. Even with the reactions switched on, Langmuirian average occupations are still found in the coexistence regions between O$^{\rm br}/-$ and O$^{\rm br}/{\rm O}^{\rm cus}$, as well as between O$^{\rm br}/-$ and CO$^{\rm br}/-$ at the lowest partial pressures. This situation changes, however, for the higher partial pressures, and particularly in the coexistence region between O$^{\rm br}/{\rm O}^{\rm cus}$ and CO$^{\rm br}/{\rm CO}^{\rm cus}$ the kinetics of the on-going surface chemical reactions leads to average site occupations, which can no longer be described by simple isotherms, and where the actual occupations exhibit complex spatial distributions and undergo large fluctuations. Since these conditions correspond also to the catalytically most active state of the surface, we will further analyze them in more detail in the section IIIB below.

Before studying the full kinetics it is useful to compare the steady-state populations with those resulting from our previous {\em ab initio} atomistic thermodynamics investigations, employing the ``constrained equilibrium'' concept \cite{reuter05a,reuter03a}. In these calculations we neglect the effect of the on-going catalytic reactions at the surface and as described in section IID above the resulting average surface populations (now identically computed by the kMC approach with the reaction processes switched off) are shown in Fig. \ref{surfpop_thermo}. When the catalytic reactions are switched on, we obtain the results shown in Fig. \ref{TOF_600}. Comparing the two plots, similarities, but also the expected noticeable differences under some environmental conditions can be discerned. The differences affect most prominently the presence of oxygen atoms at the bridge sites, where they are much more strongly bound than CO. For the thermodynamic approach only the ratio of adsorption to desorption matters, and due to the ensuing very low desorption rate, O$^{\rm br}$ is correspondingly stabilized even when there is much more CO in the gas-phase than O$_2$ (left upper part of Fig. \ref{surfpop_thermo}). The surface reactions, on the other hand, are very efficient for removing this O$^{\rm br}$ species. As result, under most CO-rich conditions in the gas-phase oxygen is faster consumed by the reactions than it can be replenished from the gas-phase. The kMC simulations, as they account for this effect, then yield a much lower average surface concentration of O$^{\rm br}$ than the thermodynamic treatment, and as a consequence show an extended stability range of surface structures with CO$^{\rm br}$ at the surface (CO$^{\rm br}/-$ and CO$^{\rm br}/{\rm CO}^{\rm cus}$ regions). It is particularly interesting to notice, that the resulting CO$^{\rm br}/-$ region, consisting of only adsorbed CO at bridge sites, does not exist in the thermodynamic phase diagram at all, cf. Fig. \ref{surfpop_thermo}. The corresponding CO$^{\rm br}/-$ ``phase'' (hatched region in Fig. \ref{TOF_600}) is thus a surface termination with defined atomic structure and composition that is entirely stabilized by the kinetics of the thermodynamically open catalytic system.

\begin{figure*}
\scalebox{0.30}{\includegraphics{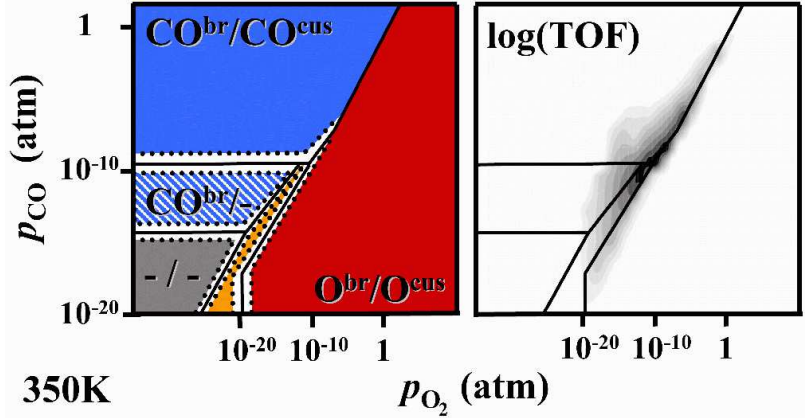}}
\caption{\label{TOF_350}
(Color online) Left panel: Steady-state surface structures obtained by first-principles kMC simulations at $T = 350$\,K (see Fig. \ref{surfpop_thermo} for an explanation of the labeling). Right panel: Map of the corresponding turn-over frequencies (TOFs) in ${\rm site}^{-1} {\rm s}^{-1}$\; ($10^{15}\; {\rm cm}^{-2}{\rm s}^{-1}$): White areas have a TOF\,$< 10^{-12}\; {\rm site}^{-1} {\rm s}^{-1} \;\;(< 10^{3}\; {\rm cm}^{-2}{\rm s}^{-1})$, and each increasing gray level represents one order of magnitude higher activity. Thus, the black region corresponds to TOFs above $10^{-5}\; {\rm site}^{-1} {\rm s}^{-1}\;\; (10^{10}\; {\rm cm}^{-2}{\rm s}^{-1})$, while in a narrow $(p_{\rm CO},p_{\rm O_2})$ region the TOFs actually peak over $10^{-3}\; {\rm site}^{-1} {\rm s}^{-1}\;\; (10^{12}\; {\rm cm}^{-2}{\rm s}^{-1})$. The plots are based on $\approx 400$ kMC simulations for different $(p_{\rm CO},p_{\rm O_2})$ conditions.}
\end{figure*}

The stability of this CO$^{\rm br}$/$-$ and the CO$^{\rm br}/{\rm CO}^{\rm cus}$ structure in the most CO-rich environments has to be considered with caution though. In such environments one would expect a reduction of the whole oxide to pure Ru metal. In our kMC simulations, the two CO-covered surface terminations persist instead, because the simulations are restricted to the kinetics involving bridge and cus surface sites. This restriction is justified under most steady-state conditions, since the creation of vacancies in deeper layers, initiated by reacting off the tightly-bound threefold coordinated lattice oxygen atoms, is extremely rare. Even if such a vacancy would be created, it would be rapidly refilled by surface oxygen atoms. However, under the most CO-rich conditions and the resulting CO-covered surface terminations (upper left part of Fig. \ref{TOF_600}) such a rapid refilling will no longer occur and the adsorbed CO molecules will eventually reduce the RuO$_2$ catalyst to pure Ru metal. Under which environmental conditions this instability of the bulk oxide against CO-induced decomposition occurs, can roughly be estimated through thermodynamic considerations \cite{reuter04a} and is indicated by the dashed line in Fig. \ref{surfpop_thermo}. As can be read off from the pressure scales in Fig. \ref{surfpop_thermo}, catalytically relevant environments are, however, quite far away from this instability limit. Although the steady-state kinetics reduces the O concentration at the surface compared to this ``constrained equilibrium'' description, we still find a significant presence of surface oxygen atoms, cf. Fig. \ref{TOF_600}, which can rapidly anneal eventually created lattice oxygen vacancies. In all catalytically relevant conditions, only a small concentration of lattice oxygen vacancies will correspondingly be present during steady-state operation, and it is the kinetics of the bridge and cus site atoms and molecules which will entirely determine the catalytic function.

Another important aspect that differs from the ``constrained thermodynamic equilibrium'' results is a more complex dependence on temperature. In the thermodynamic approach the only ruling quantities are the gas phase chemical potentials, which means that for varying temperatures equivalent surface conditions are obtained at partial pressures which scale according to eq. (\ref{mu_gas}). We can therefore e.g. summarize the stability regions of the various surface structures at 350\,K and at 600\,K in Fig. \ref{surfpop_thermo} in the same range of chemical potentials by simply indicating two different pressure scales. The only notable exception arises due to the varying configurational entropy, which leads to an increasing width of the white coexistence regions with temperature. When accounting for the kinetics, the situation becomes more complex. This is exemplified by Fig. \ref{TOF_350}, which displays the average surface populations obtained in kMC simulations at 350\,K and varying O$_2$ and CO partial pressures. The pressure range shown in Fig. \ref{TOF_350} corresponds to the same range of chemical potentials as the one in Fig. \ref{TOF_600}, but due to the kinetics of the on-going surface reactions some differences between the results at 600\,K and 350\,K can now be discerned. Scaling is still largely present though, since the change from oxygen- to CO-covered surface still occurs at roughly the same chemical potentials, indicating that the steady-state situation is mostly quite close to a ``constrained equilibrium'' \cite{reuter03a}. At the lowest partial pressures, the transitions are again Langmuir-like, although now of course with a reduced width corresponding to the lower temperature. Most important is that the surface composition with the largest fluctuations is again found for the white coexistence region between the stability region of the O$^{\rm br}/{\rm O}^{\rm cus}$ and CO$^{\rm br}/{\rm CO}^{\rm cus}$ terminations, representing also at this lower temperature the catalytically most active state.

\subsection{Catalytically active conditions}

To quantify the catalytic activity under the various conditions, the absolute TOFs were evaluated in the kMC simulations by analyzing the average occurrence of all surface reaction events. The results in the same range of partial pressures as discussed before are shown in Figs. \ref{TOF_600} and \ref{TOF_350} for 600\,K and 350\,K, respectively. As already mentioned above, the catalytic activity is at both temperatures narrowly peaked around the coexistence line between the O$^{\rm br}/{\rm O}^{\rm cus}$ and CO$^{\rm br}/{\rm CO}^{\rm cus}$ terminations, a region which we will henceforth shortly denote by the ``active region'' of the catalyst. Due to the afore discussed pressure scaling, this highest activity occurs at 600\,K at technologically more relevant pressures of the order of atmospheres, and at 350\,K accordingly at pressures typical for UHV conditions. At 350\,K the TOFs are significantly smaller than at 600\,K. For both temperatures, the calculated TOFs are in quantitative agreement with existing experimental data \cite{wang02,peden86}, both with respect to the absolute magnitude and their $(T,p_{\rm O_2},p_{\rm CO})$ dependence.

A maybe even more remarkable aspect is revealed, when analyzing the contribution of the four possible surface reaction processes to the overall reactivity. As summarized in Table \ref{tab3} these four processes have quite different energy barriers, leading to large differences in the individual rates. It is interesting to note that the energetic order among these four reactions may be rationalized on the basis of the initial bound states, since higher barriers are observed, when more strongly bound adsorbates are involved (in particular the strongly bound O$^{\rm br}$ species). On the basis of the energetic data, the lowest barrier process O$^{\rm cus}$ + CO$^{\rm br}$ involving the weaker bound surface species would have correspondingly appeared most relevant for the catalysis. Yet, in the narrowly peaked region of highest activity, its contribution is in fact only little. For instance, at $p_{\rm O_2} = 1$\,atm and $p_{\rm CO}=20$\,atm, i.e. within the most ``active region'' at $T=600$\,K, cf. Fig. \ref{TOF_600}, only 23\% of the total computed TOF of $4.6\;10^3\;{\rm site}^{-1}\;{\rm s}^{-1}$ ($4.6\;10^{18}\;{\rm cm}^{-2}\;s^{-1}$) are due to this O$^{\rm cus}$ + CO$^{\rm br}$ reaction. With 76\% it is instead the O$^{\rm cus}$ + CO$^{\rm cus}$ reaction, which dominates the overall activity, while a remaining 1\% is due to the O$^{\rm br}$ + CO$^{\rm cus}$ reaction. The highest barrier reation O$^{\rm br}$ + CO$^{\rm br} \rightarrow {\rm CO}_2$ is insignificant.

\begin{figure}
\scalebox{0.42}{\includegraphics{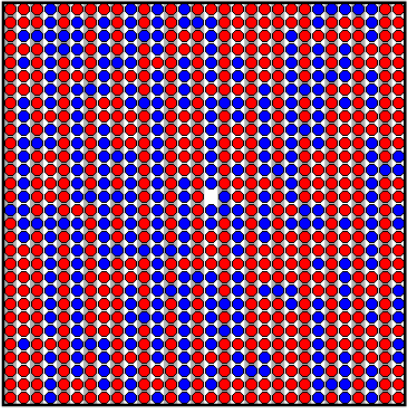}}
\caption{\label{pop600}
(Color online) Snapshot of the steady-state surface population under optimum catalytic conditions at $T = 600$\,K ($p_{\rm O_2}=1$\,atm, $p_{\rm CO}=20$\,atm). Shown is a schematic top view of the simulation area of $(30 \times 30)$ surface sites, where the substrate bridge sites are marked by gray stripes and the cus sites by white stripes. Oxygen adatoms are drawn as light gray (red) circles and adsorbed CO molecules as dark gray (blue) circles. Movies of entire kMC simulations over 500\,ns and 1\,s at various pressure and temperature conditions are also available.\cite{epaps}}
\end{figure}

Although the quantitative shares of the different reaction processes exhibit some variations, this situation holds qualitatively for all environmental conditions corresponding to the ``active region'' at both 600\,K and 350\,K: The overall TOFs are in all cases dominated by the O$^{\rm cus}$ + CO$^{\rm cus}$ reaction, although it exhibits a less favorable energy barrier compared to the O$^{\rm cus}$ + CO$^{\rm br}$ reaction. This somewhat surprising finding becomes comprehensible, when analyzing the steady-state surface populations in more detail. Fig. \ref{pop600} shows a representative snapshot at 600\,K, for peak activity pressure conditions ($p_{\rm O_2} = 1$\,atm, $p_{\rm CO}=20$\,atm). Under these conditions the kinetics builds up a surface population, in which O and CO compete for either site type at the surface. This strong competition is also reflected by the large fluctuations in the surface populations with time, apparent in Fig. \ref{evolution} displaying the time evolution of the system under exactly these gas phase conditions. At steady-state, only 11\% of all bridge sites are on average occupied by CO, while the remaining 89\% are covered by O atoms, leaving practically no vacancies at bridge sites at these high pressures. Thus, only very few CO$^{\rm br}$ are available to initiate an O$^{\rm cus}$ + CO$^{\rm br}$ reaction. On the other hand, the average CO occupation at cus sites is rather high (70\%), cf. Fig. \ref{evolution}, offering much more possibilities for the 29\% O$^{\rm cus}$ population to engage in an O$^{\rm cus}$ + CO$^{\rm cus}$ reaction. Although the low barrier O$^{\rm cus}$ + CO$^{\rm br}$ elementary process itself exhibits very suitable properties for catalysis, it can thus only occur too rarely in the full concert of all possible events to decisively affect the catalytic functionality. This emphasizes the importance of the statistical interplay and the novel level of understanding that is provided by the first-principles mesoscopic studies described here.

The just discussed results from the kMC simulations show also that explicit consideration of the evolving spatial distributions and their fluctuations are a crucial aspect for the high efficiency of catalysis. In a mean field approach the occurrence ratio of the O$^{\rm cus}$ + CO$^{\rm cus}$ to the O$^{\rm cus}$ + CO$^{\rm br}$ reaction would be determined by the ratio of the CO$^{\rm cus}$ to CO$^{\rm br}$ population available to engage with the O$^{\rm cus}$ species in a reaction, as well as the ratio of the two respective process rates. Even if such an approach would yield the same average populations at the just discussed pressure conditions (70\% CO$^{\rm cus}$ vs. 11\% CO$^{\rm br}$), a ratio of about 1:1 would still result for the two reactions, considering the 0.1\,eV difference in the two barrier heights. As mentioned before, the actually calculated ratio in the kMC simulations is, however, $\sim 3:1$. Visual inspection of the snapshot population displayed in Fig. \ref{pop600} points immediately at the reason for this difference. Despite the absence of lateral interactions among the adspecies, they are clearly not distributed in a random arrangement, but tend to cluster into small domains, which extend particularly in the direction along the bridge rows and cus trenches. This tendency arises out of the statistical interplay of all elementary processes, but has partly to do with the fact that the chemical reactions can only attack directly neighboring O and CO pairs. Adsorbed O or CO inside a formed domain can therefore only become amenable to reactions, if an efficient reshuffling of the populations occurs, e.g. through desorption events creating vacancies that are then replenished with the respective other species.

At the high pressures corresponding to the ``active region'' at 600\,K all lattice sites are almost always occupied, preventing a reshuffling due to diffusion events and suppressing particularly a reshuffling due to oxygen adsorption, which would require two neighboring vacant sites. Although the impingement rate of O$_2$ and CO from the gas phase is comparable at the peak activity conditions ($p_{\rm O_2} = 1$\,atm, $p_{\rm CO}=20$\,atm), we correspondingly observe $\sim 10^3$ more CO than oxygen adsorption events. Reshuffling at bridge sites is further inhibited due to the large desorption barrier of the strongly bound majority O$^{\rm br}$ species. All in all, this leads effectively to a much more static bridge population compared to the cus population, as reflected by the much smaller fluctuations in Fig. \ref{evolution}. This helps to stabilize certain local domain structures, which clearly differ from a random distribution. Most notably, almost all CO$^{\rm br}$ species are rapidly chaperoned by CO$^{\rm cus}$ in the left and right neighboring cus sites, cf. Fig. \ref{pop600}. They are then no longer accessible to the low barrier O$^{\rm cus} + {\rm CO}^{\rm br}$ reaction, explaining the much scarcer occurrence of this reaction compared to the $\sim 1:1$ ratio expected in a mean field picture. Since approaches based on macroscopic rate equations \cite{masel96} rely inherently on such a mean field approximation and are therefore by definition unable to properly describe the correlated fluctuations observed in our kMC simulations, it is unlikely that they could account for the correct catalytic activity in this ``active region'' of the surface, even if they were parameterized with the same first-principles data employed here. 

\begin{figure}
\scalebox{0.42}{\includegraphics{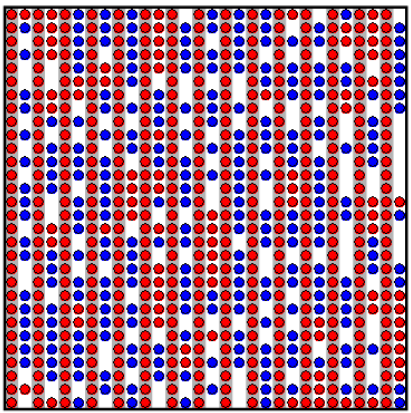}}
\caption{\label{pop350}
(Color online) Snapshot of the steady-state surface population under optimum catalytic conditions at $T = 350$\,K ($p_{\rm O_2} = 10^{-10}$\,atm, $p_{\rm CO} = 5\,10^{-10}$\,atm). Shown is a schematic top view of the simulation area of $(30 \times 30)$ surface sites, where the substrate bridge sites are marked by gray stripes and the cus sites by white stripes. Oxygen adatoms are drawn as light gray (red) circles and adsorbed CO molecules as dark gray (blue) circles. Movies of entire kMC simulations over 500\,ns and 1\,s at various pressure and temperature conditions are also available.\cite{epaps}}
\end{figure}

We arrive at essentially the same picture when analyzing the ``active region'' of the surface at $T = 350$\,K. It is, however, worthwhile to point out some specific differences. At the lower temperature, the coexistence region is much narrower due to the reduced configurational entropy, and there is correspondingly only a much smaller range of environmental conditions, which shows appreciable catalytic activity at all, cf. Fig. \ref{TOF_350}. Due to the significantly decelerated rates of the activated surface processes, the TOFs are even at optimum performance at this temperature still orders of magnitude lower than at 600\,K, actually peaking over $10^{-3}\; {\rm site}^{-1} {\rm s}^{-1}\;\; (10^{12}\; {\rm cm}^{-2}{\rm s}^{-1})$ only in a tiny $(p_{\rm CO},p_{\rm O_2})$ region. And since this optimum region corresponds now to much lower pressures, the overall average surface population is also lower. This is illustrated in Fig. \ref{pop350} by a representative snapshot at $p_{\rm O_2} = 10^{-10}$\,atm and $p_{\rm CO} = 5\;10^{-10}$\,atm, i.e. gas phase conditions for which we obtain a maximum TOF of $4.6\;10^{-3}\; {\rm site}^{-1} {\rm s}^{-1}\;\; (4.6\;10^{12}\; {\rm cm}^{-2}{\rm s}^{-1})$. While in the afore discussed ``active region'' at 600\,K there was only a 1\% vacancy concentration even at the weaker binding cus sites, this is now much higher, i.e. on average 30\% of all cus sites are empty. With 50\% CO$^{\rm cus}$ compared to 20\% O$^{\rm cus}$, there is still a higher CO than O cus population though, as was the case at the higher temperature. Similarly, O$^{\rm br}$ dominates the bridge population, however, now with virtually 100\%. Correspondingly, the lowest barrier O$^{\rm cus} + {\rm CO}^{\rm br}$ reaction, which would require the presence of CO$^{\rm br}$ species, is almost completely suppressed at this temperature, amounting only to a mere 0.4\% of the total TOF. 

The overall population at 350\,K is therefore much closer to the one expected within the ``constrained thermodynamic equilibrium'' approach, cf. Fig. \ref{surfpop_thermo}. This is due to the fact that the scarce reaction events occurring at this temperature lead only to a modest reduction of the O$^{\rm cus}$ population. The surface structure and composition becomes therewith also more similar to some of the terminations that have hitherto been stabilized and atomically characterized in surface science UHV experiments. \cite{fan01,wang02,over00,kim01a,wang01,kim01b,seitsonen01,kim03,paulus03} The problem with such studies is that depending on the specific preparation recipe quite different surface structures can be kinetically frozen in. These need then not to have anything to do with the optimum disordered and dynamic adsorbate composition, built up by the intricate interplay of all surface processes under steady-state conditions. While cautious interpretation of corresponding experimental data does in principle provide a wealth of atomic-scale information, direct conclusions concerning the high-pressure steady-state reactivity are problematic. This holds in particular for reactivity analyses e.g. through temperature programmed reaction spectroscopy. If they are based on initial surface populations that are qualitatively different from the real steady-state ensemble, qualitatively wrong conclusions can be deduced \cite{wendt03,wendt04}.

\subsection{Uncertainties and compensation effect}

With the advent of first-principles TOF maps as those shown in Figs. \ref{TOF_600} and \ref{TOF_350}, a more detailed data base is becoming available for comparison with experiments, which will eventually advance our microscopic understanding of catalysis. In order to fully appreciate the quality gained by the underlying first-principles statistical mechanics, it is, however, mandatory to qualify its accuracy. Since several scales are bridged, uncertainties can arise at many levels. In the electronic regime, we have errors in the energetics both due to approximations in the numerical setup and the exchange-correlation treatment. On the way to the mesoscale, it is the use of transition-state theory. Finally, at the statistical mechanics level, it is the suitability of the employed lattice model of the surface and the set of elementary processes occurring on it (e.g. the neglect of metal precipitates, vacancies or steps). 

While it seems natural that some degree of coarse-graining on the way to macroscopic functionality should be permissible and would thereby alleviate some of the errors introduced in the electronic-level description, this must still be carefully checked for every system. In the present example of CO oxidation at RuO$_2$(110) we are confident that the employed lattice and process list captures all of the essential physics of the active surface at steady-state discussed here. This confidence stems from extensive theoretical and experimental work over the last years, which established the prominent role of the two adsorption sites, and confirmed in particular, that although steps and domain boundaries between RuO$_2$ grains are present, their influence is not significant. \cite{reuter02a,reuter03a,over03,boettcher99,kim01a,wang01,kim01b,seitsonen01,kim03,paulus03,he05} Our error analysis concentrates therefore on the individual process rates entering the statistical mechanical simulations. To account for the effective uncertainty introduced by the approximate prefactors and activation barriers, we correspondingly reran the kMC simulations for rates that were systematically varied by a factor of 100, which would e.g. correspond to a barrier variation of $\pm 0.2$\,eV at 600\,K. It is important to note, that these variations were in all cases done in such a way that detailed balance, cf. eq. (\ref{equilibrium2}), was not violated. If, for example, an adsorption rate was lowered (reflecting e.g. an increased activation barrier), so was the connected desorption rate, since the particle would have to overcome an equally increased barrier.

As a first observation, our results are basically not affected by the uncertainty in the eight diffusion processes. Due to the mostly quite high barriers, most of these events have an extremely low process rate anyway. Under most environmental conditions and especially in the catalytically ``active region'' their occurrence is further suppressed due to the almost 100\% occupancy of all lattice sites at all times. Correspondingly, the number of executed diffusion events is so low, that we could in fact completely switch them off in the kMC simulations, without observing a significant variation in our results. Our quantitative simulations can therefore not support the speculations on a specific role of the O$^{\rm cus}$ species primarily in replenishing once formed bridge vacancies.\cite{wendt03,wendt04} Under high-pressure steady-state conditions, such diffusion events can basically not occur, highlighting again that an incautious extrapolation of interpretations derived from UHV experiments can be quite misleading.

As for the remaining 18 processes, or better 9 rates plus detailed balance, the central result of our error analyses is that the central properties discussed here, e.g. the overall structure of the steady-state surface populations and the position of optimum catalytic efficiency in $(T,p_{\rm O_2},p_{\rm CO})$ space, are surprisingly insensitive to variations in the individual rates. This is mostly due to the fact that these properties are not determined by one singular process alone, but by the action of many players. This puts immediately more emphasis on the relative rate {\em differences} between the various contributing processes instead of the absolute values, and systematic errors in the rates like e.g. a general overbinding of a given exchange-correlation functional are then less consequential. A too high desorption barrier goes then e.g. hand in hand with a too high reaction barrier, so that the reduced rate for one reaction process is partially compensated by an increased number of available surface species, leaving the net number of executed reactions relatively unaffected. It is important to realize that this implies the necessity of a consistent treatment of all participating processes, and a combination of different calculations (employing different approximations) or of theoretical and experimental parameters could have spoiled the description. Interestingly, a recent study by Honkala {\em et al.} arrived at quite similar conclusions, when studying ammonia synthesis at a ruthenium nanoparticle with a comparable first-principles statistical mechanics approach.\cite{honkala05}

We therefore find that the variations of the rates led at most to shifts of up to 2 orders of magnitude in transition pressures from one surface structure to the other, which does not affect the overall structure of the steady-state population plots in Fig. \ref{TOF_600} and \ref{TOF_350} at all. Similarly untouched is the intimate connection between the narrowly peaked region of high catalytic activity and the realization of a disordered dynamic surface ``phase'', where the kinetics builds a surface composition in which O and CO occupy both site types. Under these conditions the catalytic activity results from the concerted action of many processes, in fact always at least 9 out of our total of 26 processes are executed with significant frequencies. This strong interplay makes the resulting optimum mix at the catalyst surface highly adaptive and renders the corresponding maximum TOFs in the core region of highest catalytic activity quite insensitive to modest errors in the rates. Even when one or more individual process rates were varied by a factor of 100, this led only to changes in the peak TOF by a factor of 4. When one reaction barrier was e.g. increased, either the spatial distribution of the surface populations adapted to yield a quite unchanged contribution of the corresponding process to the total TOF, or one or several of the other reaction processes occurred more often and led in this way to a virtually unchanged net TOF. 

These results are at variance with the frequently employed concept of one single rate-determining step, introduced initially to simplify the analysis of a reaction network.\cite{boudart68} Under the conditions of high catalytic activity, there is in the present system no high degree of rate control \cite{dumesic91,campbell01} by one specific process, since the system has at any time several alternatives along which to proceed. We believe that this is a quite generic feature of catalytic systems. Only a first-principles statistical mechanics study, as the one described above, reveals these effects. Crucial is the consideration of a high number of elementary processes (compared to typical rate equation approaches focusing on few generic representatives of process types) {\em and} the consideration of the explicit correlated spatial distribution at the surface (compared to the mean field approximation underlying rate equation approaches). Although perhaps unexpected \cite{campbell04}, the abundance of the CO$^{\rm cus} - {\rm O}^{\rm cus}$ nearest neighbor pairs (as well as CO$^{\rm br} - {\rm O}^{\rm cus}$ nearest neighbor pairs) is then apparently playing a role of similar importance as the energy barriers of individual processes. In this sense, the frequently requested {\em chemical accuracy} for the description of individual processes appears to be a misleading concept. At least for the present system, a careful combination of DFT and statistical mechanics is at least as important as good data for the energy barriers.

\section{Summary}

In summary we have presented a first-principles statistical mechanics approach to quantitatively study the steady-state situation of heterogeneous catalysis. Density-functional theory is employed together with transition state theory to accurately obtain the energetics of all relevant elementary processes, subsequently solving the statistical mechanics problem by kinetic Monte Carlo simulations. The key merits of this two-step approach are that microscopic insight into the system can be gained by following its full dynamics from picoseconds up to seconds, while simultaneously accounting for all atomic-scale correlations, fluctuations, and spatial distributions at the catalyst surface.

In its application to the CO oxidation at the RuO$_2$(110) model catalyst, turnover frequencies in unprecedented agreement with existing experimental data are achieved in a wide range of $(T,p_{\rm O2},p_{\rm CO})$ conditions ranging from UHV up to technologically relevant pressures of the order of atmospheres and up to elevated temperatures. The catalytic activity is narrowly peaked at $(T,p)$ conditions, where the surface kinetics builds a disordered and dynamic adsorbate composition with O and CO competing for both prominent cus and bridge adsorption sites at the surface. The statistical analysis of the surface dynamics and of the various elementary processes in steady-state operation reveals several surprising results. Most notably, the chemical reaction with the most favorable energy barrier contributes only little to the overall CO$_2$ production at maximum activity conditions. Although the process itself features very suitable properties for catalysis, it can only occur too rarely in the full concert of all possible events to decisively affect the observable macroscopic functionality. This strong interplay of a larger number of elementary processes makes the resulting optimum mix at the catalyst surface highly adaptive and renders the corresponding maximum TOFs in the core region of highest catalytic activity quite insensitive to modest errors in the rates. This challenges established concepts like the idea of one rate-determining step or the necessity of chemical accuracy for the description of individual processes, and clearly illustrates the new quality and novel insights gained by the modern first-principles statistical mechanics methodology developed and applied here.

\section{Acknowledgements}

This work was partially supported by the Deutsche Forschungsgemeinschaft (DFG) in the priority program SPP-1091, and by the EU under contract no. NMP3-CT-2003-505670 (NANO$_2$). KR gratefully acknowledges enlightening discussions with Daan Frenkel and Mark Miller during his DFG-financed stay at the FOM Institute for Atomic and Molecular Physics (AMOLF).

\end{document}